\documentclass{article}

\usepackage[nonatbib,final]{neurips_2024}
\usepackage[utf8]{inputenc} %
\usepackage[T1]{fontenc}    %

\usepackage[
backend=biber,
style=numeric-comp,
natbib=true,
maxbibnames=99
]{biblatex}

\AtEveryBibitem{
    \clearfield{urlyear}
    \clearfield{urlmonth}
}

\renewbibmacro*{doi+eprint+url}{
\iftoggle{bbx:url}     {\iffieldundef{doi}{\usebibmacro{url+urldate}}{}}     {}
\newunit\newblock   \iftoggle{bbx:eprint}     {\usebibmacro{eprint}}     {}   \newunit\newblock   \iftoggle{bbx:doi}     {\printfield{doi}}     {}}

\PassOptionsToPackage{hyphens}{url}
\usepackage{hyperref}      %
\usepackage{url}            %
\usepackage{booktabs}       %
\usepackage{amsfonts}       %
\usepackage{nicefrac}       %
\usepackage{microtype}      %
\usepackage{xcolor}         %
\usepackage{graphicx}
\usepackage{fontawesome5}
\usepackage{array}
\usepackage{colortbl}
\definecolor{head}{HTML}{d0e0e3}
\usepackage{wrapfig}
\usepackage{enumitem}
\usepackage{longtable}
\usepackage[size=normalsize]{caption}
\usepackage{cleveref}
\usepackage{placeins}

\setenumerate{itemsep=6pt,topsep=6pt}

\title{Responsible Reporting for Frontier AI Development}

\author{%
  \hspace{-3mm}\parbox{\linewidth}{\centering\bfseries%
  Noam Kolt,\!$^{1}$\hspace{0.3pt}\thanks{Work done at Google DeepMind.}\enskip \thanks{Correspondence to \href{mailto:noam.kolt@mail.utoronto.ca}{\texttt{noam.kolt@mail.utoronto.ca}}.}\;\, Markus Anderljung,\!$^{2}$\, Joslyn Barnhart,\!$^{3}$\, Asher Brass,\!$^{4}$\\ Kevin Esvelt,\!$^{5}$\, Gillian K. Hadfield,\!$^{1, 6}$\, Lennart Heim,\!$^{2}$\, Mikel Rodriguez,\!$^{3}$\\ Jonas B. Sandbrink,\!\hspace{-1pt}$^{7}$\, Thomas Woodside$^{8}$
  \parbox{\linewidth}{\vspace{1em}\mdseries\centering
  $^{1}$University of Toronto,
  $^{2}$Centre for the Governance of AI,
  $^{3}$Google DeepMind,\\
  $^{4}$Institute for AI Policy and Strategy,
  $^{5}$Massachusetts Institute of Technology,\\ 
  $^{6}$Vector Institute for AI,   $^{7}$University of Oxford,\\
  $^{8}$Center for Security and Emerging Technology
  }
}
}

\begin{document}

\maketitle

\begin{abstract}
  Mitigating the risks from frontier AI systems requires up-to-date and reliable information about those systems. Organizations that develop and deploy frontier systems have significant access to such information. By reporting safety-critical information to actors in government, industry, and civil society, these organizations could improve visibility into new and emerging risks posed by frontier systems. Equipped with this information, developers could make better informed decisions on risk management, while policymakers could design more targeted and robust regulatory infrastructure. We outline the key features of responsible reporting and propose mechanisms for implementing them in practice.
\end{abstract}

\begin{figure}[hb!]
    \centering
    \includegraphics%
    {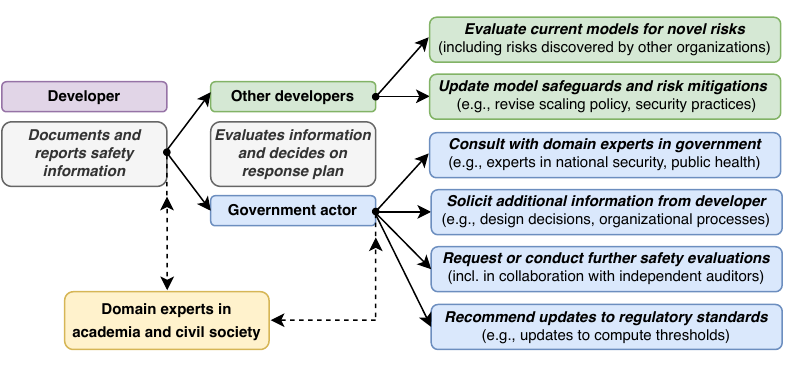}
    \vspace{3mm}\caption{\textbf{A framework for responsible reporting}. Developers disclose safety-critical information to government actors and other developers, which decide on appropriate technical, organizational, and policy responses. Independent domain experts in academia and civil society receive key information and provide guidance to both developers and government actors.}
    \label{fig:1}
\end{figure}

\newpage

\section{Introduction}\label{sec:1-introduction}

Information is the lifeblood of good governance \citep{stephenson2011,
vanloo2019}. Effectively responding to the new and emerging risks
presented by frontier AI systems \citep{anderljung2023, zotero-3420}
requires up-to-date and reliable information about those systems and
their impact on society \citep{birhane2024, shevlane2023,
weidinger2023}. There is growing consensus among experts in AI safety
and governance that reporting safety information to trusted actors in
government and industry is key to achieving this goal
\citep{bommasani2023, kapoor, schuett2023}. This is particularly the
case for frontier models, i.e., highly capable foundation models that
could pose severe risks to public safety \citep{anderljung2023,
phuong2024, fitch2024}.

Early efforts to facilitate reporting safety information made important
strides. The AI Incident Database established by the Partnership on AI
contains more than 2,000 reports of AI harms \citep{mcgregor2022,
mcgregor2021}. The database, however, is limited to tangible
harms caused by deployed AI systems, as is the case for related
initiatives \citep{turri2023}. These databases do not track
anticipated risks, vulnerabilities, or near-misses
\citep{hoffmannmia2023, shrishak2023}, and dedicate comparatively
little attention to larger-scale or catastrophic risks
\citep{bengio2023, ukgovernment2023, hendrycks2023}.

But the tide is changing. Recognizing the growing need to share
information about AI safety with government actors, several leading
developers committed to the U.S. government to ``reporting their AI
systems' capabilities, limitations, and areas of appropriate and
inappropriate use'' and undertook to engage in ``third-party discovery
and reporting of vulnerabilities in their AI systems''
\citep{house2023}. Some developers made additional commitments to
share information with companies and governments
\citep{frontiermodelforum}, including to provide actors in the UK
government with ``early or priority access to models for research and
safety purposes to help build better evaluations''
\citep{departmentforscienceinnovationandtechnology2023a}. The UK
government has also requested access to, and published, details
concerning the safety practices of several leading AI companies
\citep{departmentforscienceinnovationandtechnology2023}.

National governments and international institutions are also taking
concrete steps to implement AI safety reporting. The European Union's AI
Act imposes stringent reporting obligations on the providers of
high-risk AI systems \citep{europeanparliamenta}. An executive order
issued by President Biden requires that AI developers provide the U.S.
federal government with information regarding ``activities related to
training, developing, or producing dual-use foundation models,'' as well
as information regarding ``the ownership and possession of the model
weights'' and the results of ``red-team testing'' \citep{thewhitehouseExecutiveOrderSafe2023}. These requirements
would initially apply to any model trained using more than
$10^{26}$ operations, or any model using primarily genetic
sequence data trained using $10^{23}$ operations. The OECD, meanwhile, has
convened an expert group to develop an AI incident reporting framework
\citep{oecd.ainetworkofexperts}.

Given this increasingly complex institutional context, distilling the
key features of AI safety reporting is especially important. We aim to
make headway on this challenge by clarifying the goals of reporting
safety-related information (\Cref{sec:2-goals-of-reporting}), describing the content of
this information and to which actors it could be reported (\Cref{sec:3-decision-relevant-information}), proposing institutional mechanisms to facilitate reporting (\Cref{sec:4-institutional-framework}), and tackling potential hurdles to implementing these
mechanisms in practice (\Cref{sec:5-implementation}). Taken together, these
contributions complement and provide guidance for more concrete efforts
to establish reporting frameworks, including multiple concurrent efforts
being undertaken by actors in government, industry, and civil society.

\section{Goals of reporting}\label{sec:2-goals-of-reporting}

As in other industries with long-standing reporting practices, including
healthcare, finance, and aviation \citep{mcgregor2021, raji2022,
turri2023}, information disclosures aim to achieve several goals. In
the case of frontier AI systems, we focus on three main goals of
reporting: (1) raising awareness among key stakeholders with regard to
societal-scale impacts and risks from AI technologies; (2) incentivizing
AI developers to adopt more robust risk management and safety practices;
and (3) increasing regulatory visibility to enable policymakers to
effectively respond to new risks, especially risks which government
actors are best positioned to address.

\begin{figure}[ht!]
    \centering
    \includegraphics%
    [scale=1.1]{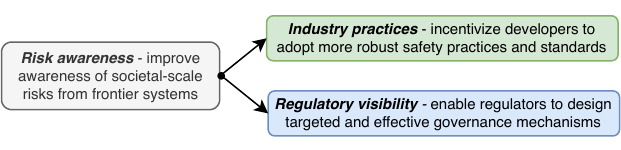}
    \caption{Goals of reporting safety information}
    \label{fig:2}
    \vspace{-12pt}
\end{figure}

\subsection{Risk awareness}\label{risk-awareness}

In its simplest form, the case for reporting information about AI risks
and vulnerabilities can be summarized as follows: ''To make AI safer, we
need to know when and how it fails'' \citep{arnold2021}. Access to
such information is especially crucial in the case of frontier AI
systems, whose risk profiles are continually changing due to their often
unpredictable capabilities \citep{ganguli2022, phuong2024, wei2022,
weidinger2023, fitch2024}.

Across both industry and government, ``reporting builds a norm of
admitting mistakes, noticing them, and sharing lessons learned''
\citep{manheim2023}. Information about safety incidents and failures
can offer valuable lessons on how and where risk mitigation measures
fail \citep{mcgregor2021}. Alongside improving general awareness of
these risks, both developers and government actors, supported by
independent domain experts, need a detailed understanding of these risks
and the methods for mitigating them. We focus on each in turn.

\subsection{Industry practices}\label{industry-practices}

Sharing information with other developers about risks from frontier AI
systems and the measures for addressing them enables developers to
design better risk management strategies and safety practices~\citep{avin2021}. For example, a developer may update its safety
levels \citep{anthropic2023}, preparedness framework
\citep{openai2023}, or capabilities scaling policy
\citep{deepmind2023} in response to information about new capabilities
or risks \citep{anderson-samways2024,
modelevaluationandthreatresearch2023}. Information collected via
reporting could also assist developers in establishing emergency
response plans, internal audit procedures, and customer screening
processes \citep{egan2023, schuett2023}.

In addition to informing particular risk management practices, reporting
could foster a stronger safety culture across the AI industry
\citep{manheim2023, ojewale2024} and bring it closer in line with
well-established reporting practices and norms in other industries, such
as in the aviation industry \citep{federalaviationadministration2023,
madsen2015}. For instance, regulation that mandates reporting the
results of AI safety evaluations
\citep{thewhitehouseExecutiveOrderSafe2023} could deter developers
from accelerating development at the expense of safety
\citep{askell2019, dafoe2023}. By enabling external actors and domain
experts to verify the claims made by AI developers about the safety of
their systems \citep{avin2021, brundage2020, sastry2024}, reporting
would subject frontier AI developers and their products to increased
scrutiny. Consequently, less cautious actors would be incentivized to
invest more in safety and adopt best practices employed by other
organizations. As in relatively mature industries such as healthcare and
finance, risk mitigation and safety could become an inherent and
uncontroversial part of frontier AI development and deployment
\citep{dobbe2022, hendrycks2022}.

\subsection{Regulatory visibility}\label{regulatory-visibility}

Reporting information about AI risks and potential mitigations is
critical to informing the priorities and actions of policymakers.
Without meaningful visibility into the technology's design and use,
policymakers cannot determine appropriate regulatory objectives, let
alone build appropriate regulatory infrastructure
\citep{anderljung2023, hadfield2023}. Accurate and up-to-date
information about frontier AI systems and their impact is key to
enabling policymakers to address the risks posed by these systems.

A reporting framework designed to furnish policymakers with
safety-critical information will help address these concerns. Equipped
with reliable and timely information about frontier AI systems,
policymakers will be able to make better informed decisions about the
goals and methods of regulation, and acquire the resources needed to
take appropriate action \citep{clark2023, whittlestone2021}. For
example, policymakers will be able to design or implement standards that
are more responsive to trends in AI development and, ideally, preempt
nascent and emerging risks \citep{arbel2023, kaminski2023,
kolt2023}. In particular, information collected from reporting will
assist policymakers in tackling risks that government actors are best
positioned to address, as observed by the UK government's AI Safety
Institute
\citep{secretaryofstateforscienceinnovationandtechnology2023}. For
instance, upon receiving a report concerning national security risks
(e.g., new cyber capabilities or biological capabilities), national
security experts in government could propose additional model
evaluations or governance measures.

\section{Decision-relevant
information}\label{sec:3-decision-relevant-information}

The following section describes the categories of information that
developers could report in the proposed framework, as well as the
recipients of this information. The categories in \Cref{tab:1} -
\emph{development and deployment}, \emph{risks and harms}, and
\emph{mitigations} - are designed to provide government actors,
developers, and independent domain experts with information that will
assist in deciding on appropriate technical, organizational, and policy
responses to novel AI capabilities and risks. In addition, the
categories broadly align with recent regulatory regimes, including the
disclosure requirements in the U.S. executive order
\citep{thewhitehouseExecutiveOrderSafe2023}, the EU AI Act
\citep{europeanparliamenta}, and the UK proposal for AI risk reporting
\citep{departmentforscienceinnovationandtechnology2023b}.

\begin{table}[htb]
\caption{Information categories, content, and recipients}
\label{tab:1}
\centering
\arrayrulecolor{head!30!black}
\renewcommand{\arraystretch}{1.5}
\begin{tabular}{|p{0.19\linewidth}|p{0.735\linewidth}|}\hline
\rowcolor{head}\bfseries\centering\arraybackslash Category & \bfseries\centering\arraybackslash Content \\\hline
\begin{tabular}[t]{@{}>{\centering\arraybackslash}p{\linewidth}@{}}\bfseries Development and deployment\\[0.76cm] {\large\faUsers}\enskip\bfseries Recipients:\\[-5pt] Government actors\end{tabular} & \begin{tabular}[t]{@{}p{\linewidth}@{}}\textbf{Details of state-of-the-art systems} - copies of publicly available technical reports, including system and model cards, and additional information on training techniques, resources, and model capabilities.\\ \textbf{Information on current and upcoming training runs} - description of architecture, compute, data collection, curation, filtering, and human feedback, training objectives (e.g., reward functions), and training techniques.\\
\textbf{Current and anticipated applications} - description of the domains in which a model is currently deployed or anticipated to be deployed, the range of tasks they perform or are anticipated to perform, and usage trends and statistics.\end{tabular} \\\hline
\begin{tabular}[t]{@{}>{\centering\arraybackslash}p{\linewidth}@{}}\textbf{Risks and harms}\\[0.38cm] {\large\faUsers}\enskip\bfseries Recipients:\\[-5pt] \mbox{Government actors}, developers, and independent domain experts\end{tabular} & \begin{tabular}[t]{@{}p{\linewidth}@{}}\textbf{Pre-deployment and post-deployment risk assessments} - results of internal and external safety evaluations, including results of red-teaming and bounty programs.\\ 
\textbf{Concrete harms and safety incidents} - description of incidents in which a system caused death or serious injury, damage to critical infrastructure, environmental harm, cybersecurity incidents, or other concrete harms, as well as harms that did not materialize (“near misses”).\\ \textbf{Dual-use and dangerous capabilities} - evidence of a system exhibiting the ability to perform deception or manipulation, dual-use cyber capabilities or biological capabilities, weapons development, indications of the ability to engage in long-term planning, power-seeking, or other dangerous capabilities.\end{tabular} \\\hline
\begin{tabular}[t]{@{}>{\centering\arraybackslash}p{\linewidth}@{}}\textbf{Mitigations}\\[0.38cm] {\large\faUsers}\enskip\bfseries Recipients:\\[-5pt] \mbox{Government actors}, developers, and independent domain experts\end{tabular} & \begin{tabular}[t]{@{}p{\linewidth}@{}}\textbf{Model alignment and safeguards} - detailed explanation of alignment techniques, steps taken to prevent malicious use and other misuse (e.g., out-of-domain use), safety evaluations, and monitoring procedures.\\ 
\textbf{Organizational risk management} - description of security standards, personnel and customer screening, auditing procedures, review processes, or other internal governance mechanisms, including circumstances in which such procedures were not effective or were not adopted.\end{tabular}\\\hline
\end{tabular}
\end{table}

Furnishing policymakers and domain experts with the above information is
key to overcoming the inherent information deficit between industry and
government \citep{karkkainen2008}. Despite repeated calls to provide
governments with more comprehensive and consequential information
relating to frontier AI technologies \citep{avin2021,
whittlestone2021}, regulators have often been caught off-guard, as
exemplified by early drafts of the EU AI Act altogether failing to
address foundation models.

As in other domains \citep{vanloo2019}, government actors need a deep
understanding of the underlying technology, the resources required to
build it, and the risks it may pose \citep{anderljung2023}. For
example, aviation regulators are authorized to conduct sweeping
inspections of new aircraft technologies (e.g.,
\citep{federalaviationadministration2023a}), while financial
regulators have privileged access to cutting-edge financial products and
services in order to assess their anticipated impact on consumers and
markets (e.g., \citep{cfpb2022}). Without comparable information on
frontier AI systems (including information concerning development,
risks, and mitigations), policymakers and domain experts will be unable
to assess for themselves the systems' risk profiles or decide on the
appropriate governance sites and mechanisms \citep{kolt2023,
kolt2024}.

\begin{wrapfigure}{r}{0.35\textwidth}
    \vspace{-1.25mm}
    \centering
    \includegraphics[width=0.98\linewidth]
    {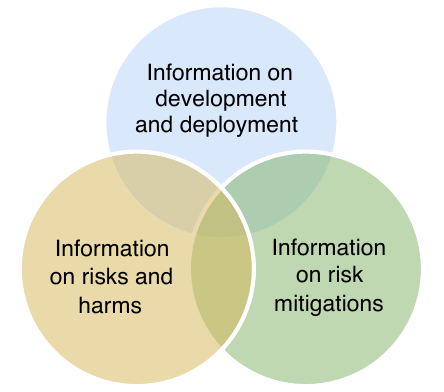}
    \vspace{3.792mm}
    \caption{There is significant overlap between information pertaining to development and deployment and information pertaining to risks, harms, and mitigations.}
    \label{fig:3}
\end{wrapfigure}
Importantly, there is likely to be significant overlap between
information pertaining to development and deployment (disclosed only to
government actors) and information pertaining to risks and risk
mitigations (disclosed to government actors, developers, and independent
domain experts). Disentangling these two categories is not
straightforward. For example, information regarding state-of-the art
alignment techniques is both a model capability as well as a risk
mitigation tool, making it unclear whether, or to what extent, such
information should be disclosed to developers and independent domain
experts, or only to government actors.

While we do not propose a precise definition distinguishing between the
different categories, the appendices offer a concrete illustration of
the kind of information that could fall into each category. \Cref{App:A-Cybersecurity,App:B-Biosecurity}, which relate to cybersecurity and biosecurity, respectively, help
shed light on which information would be disclosed to government actors
only and which information would be disclosed to government actors,
developers, and independent domain
experts.

\section{Institutional framework}\label{sec:4-institutional-framework}

\subsection{Contributors}\label{contributors}

Many different organizations are involved in developing and deploying
frontier AI models \citep{epoch2024}. These include organizations in
industry, academia, and civil society, and across different geographies.
While some organizations make their models accessible only via API
(e.g., OpenAI, Anthropic), others publicly release the model weights
subject to software licenses (e.g., Meta, EleutherAI).

Given that each of these organizations has expertise developing
different models and deploying them in different contexts, each
organization could offer valuable safety information. For example, an
organization with extensive red-teaming experience could assist other
organizations in designing protocols for external scrutiny of models
\citep{anderljung2023a}. Meanwhile, an organization that has developed
methods to mitigate malicious use of its models could share those
methods with other organizations.

Subject to the implementation challenges addressed below (\Cref{sec:5-implementation}), we
suggest that all developers of frontier models could participate in the
proposed framework and that the information they contribute could
improve visibility into AI risks and mitigations.

\subsection{Recipients}\label{recipients}

\emph{\textbf{Government actors}.} As in disclosure regimes in other
domains \citep{mcgregor2021, raji2022, turri2023}, government actors
are important recipients of the information provided under the proposed
framework. Key characteristics for government actors include the
following:

\begin{enumerate}

\item
  \textbf{Information security} - capacity to protect highly sensitive
  information and prevent its proliferation or misuse.
\item
  \textbf{Technical competence} - ability to understand, analyze, and
  draw conclusions from the information reported, including relevant
  domain expertise.
\item
  \textbf{Governance capacity} - organizational resources and legal
  authority to design and implement policy responses.
\item
  \textbf{Independence} - incentives and motivation to systematically
  and impartially execute policy responses.
\end{enumerate}

In the United States, key actors include the U.S. Artificial
Intelligence Safety Institute \citep{u.s.departmentofcommerce2023,
u.s.departmentofcommerce2024}, established through the National
Institute of Standards and Technology (NIST), an agency that itself has
significant in-house technical expertise and published an AI Risk
Management Framework
\citep{nationalinstituteofstandardsandtechnology2021}. Another key
actor is the White House Office of Science and Technology Policy (OSTP), which
released a Blueprint for an AI Bill of Rights
\citep{thewhitehouse2022} and has been involved in facilitating model
evaluations and securing voluntary commitments from leading AI
developers \citep{thewhitehouse2023, house2023a, house2023}. Other
relevant actors include the National Security Council (NSC), Bureau of
Industry and Security (BIS), Federal Trade Commission (FTC), and
possibly new government bodies. In the United Kingdom, key actors
include the UK government's AI Safety Institute, which has indicated
that it will work on conducting evaluations of advanced AI systems and
facilitating information exchange
\citep{departmentforscienceinnovation&technology23}.

Importantly, the combination of above characteristics is a new `muscle'
that governments will need to grow and flex. Effective reporting
requires broad technical and sociotechnical capacity-building, which
will require significant time and talent. In addition, it is worth
noting that different government bodies might be better positioned to
receive different types of information, instead of a single government
body receiving all information disclosed under the framework. For
example, dedicated cybersecurity agencies and biosecurity agencies might
be the preferred recipients of information in their respective domains.\vspace{1em}

\begin{figure}[htb]
    \centering
    \includegraphics[scale=1.1]{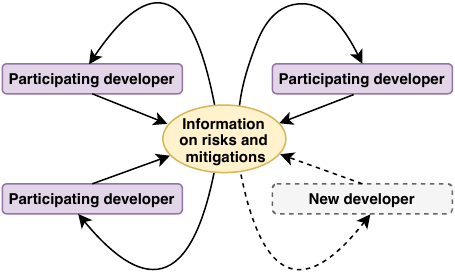}
    \vspace{0.75em}
    \caption{The principle of reciprocity incentivizes developers
to join the responsible reporting framework by providing them with
useful safety information.}
    \label{fig:4}
\end{figure}

\emph{\textbf{Developer reciprocity.}} As to which developers receive
information under the reporting framework, we propose a principle of
reciprocity according to which only developers that \emph{contribute}
information under the proposed framework will \emph{receive} information
under the framework. This principle both incentivizes developers to
participate in the framework and prevents non-participating developers
from free-riding. As illustrated in \Cref{fig:4} and \Cref{tab:1}, participating
developers will only receive information relating to risks, harms, and
mitigations, not commercially sensitive information relating to model
development and deployment - which will be disclosed to government
actors only.
\medskip

\medskip

\emph{\textbf{Independent domain experts}.} As illustrated in \Cref{fig:1}, independent domain experts in academia and civil society play
two key roles in the proposed framework. First, domain experts will
receive safety-critical information comparable to the information
provided to participating developers. Second, domain experts will
provide guidance to government actors with regard to the categories and
content of information that developers report within the framework. In
addition to the capacity to maintain highly effective information
security, participating domain experts should have the following
characteristics:

\begin{enumerate}

\item
  \textbf{Domain-specific expertise} - significant expertise in a
  risk-relevant field, especially a field or subfield in which
  government actors lack sufficient expertise.
\item
  \textbf{Security mindset} - ability and inclination to identify
  vulnerabilities, risks, and potential exploitations of AI systems and
  the environments in which they operate.
\end{enumerate}

\vfill\null
\newpage 

\subsection{Documentation and
disclosure}\label{documentation-and-disclosure}

\emph{\textbf{Documentation}.} While there does not currently exist a
dedicated government database for information relating to AI development
and deployment \citep{clark2023, turri2023}, there are many
mechanisms for documenting such information. These include data sheets
\citep{gebru2021}, model cards \citep{mitchell2019},
reward reports \citep{gilbert2023}, system cards (e.g.,
\citep{openai2023a, openai2023b}), model reports
\citep{geminiteam2023, reid2024, touvron2023}, and ecosystem graphs
\citep{bommasani2023a,bommasani}. Documentation may also need to be tailored to
the particular model or system \citep{norval2022}. For example,
meaningful insight into models using reinforcement learning from human
feedback (RLHF) can only be achieved via granular information on the
relevant human feedback, reward model, and policy \citep{casper2023,
lambert2023}.

Drawing on these and other documentation practices, we offer a
preliminary picture of how developers could record and communicate
information under the proposed framework - as illustrated in \Cref{tab:2}.
Notably, leading developers already report some of this information
(including the developers of GPT-4, Claude 3, Llama 2, and Gemini).

\begin{table}[htb]
\caption{Developer documentation for each category of safety
information}
\label{tab:2}
\centering
\arrayrulecolor{head!30!black}
\renewcommand{\arraystretch}{1.5}
\begin{tabular}{|p{0.19\linewidth}|p{0.735\linewidth}|}\hline
\rowcolor{head}\bfseries\centering\arraybackslash Information categories & \bfseries\centering\arraybackslash \raisebox{-0.5\baselineskip}{Documentation} \\\hline
\begin{tabular}[c]{@{}>{\centering\arraybackslash}p{\linewidth}@{}}\bfseries Development and deployment \end{tabular} & \begin{tabular}[c]{@{}p{\linewidth}@{}}Model cards for current models and models under development, including details of data collection, model training and finetuning, evaluation metrics, intended uses, ecosystem dependencies (e.g., compute sponsors, API access), and model maintenance plan.
\end{tabular} \\\hline
\begin{tabular}[c]{@{}>{\centering\arraybackslash}p{\linewidth}@{}}\textbf{Risks and harms}\end{tabular} & \begin{tabular}[c]{@{}p{\linewidth}@{}}Results of pre-deployment and post-deployment safety evaluations (including underlying code and data), records of all safety incidents (including actors involved, causes of incidents, and responses to incident), and additional threat intelligence and impact assessments.
\end{tabular} \\\hline
\begin{tabular}[c]{@{}>{\centering\arraybackslash}p{\linewidth}@{}}\textbf{Mitigations}\end{tabular} & \begin{tabular}[c]{@{}p{\linewidth}@{}}Description of technical safeguards for preventing misuse and accident risks (including alignment and monitoring procedures) and organizational measures (e.g., documents establishing internal safety and governance structures).\end{tabular}\\\hline
\end{tabular}
\end{table}

Although major jurisdictions increasingly mandate reporting, the precise
scope and form of reporting remain open to interpretation. For example,
the U.S. executive order requires that developers provide ``information,
reports, or records'' regarding ``training, developing, or producing
dual-use foundation models'' and ``the ownership and possession of the
model weights'', but does not specify precisely what information those
reports must contain \citep{thewhitehouseExecutiveOrderSafe2023}. Nor
does the executive order prescribe the exact form in which information
should be documented, leaving these fine-grained, yet important, tasks
to future policy instruments.

In addition, given that developers' safety evaluations and reporting
practices are likely to evolve over time, policymakers will need to
continually refine their governance responses to information received,
which will require ongoing investment in both technical competence and
institutional capacity (see~\Cref{recipients}).

\emph{\textbf{Disclosure process}.} In addition to carefully documenting
safety information, developers will need to securely communicate that
information to the intended recipients. Doing so is critical to
preventing the information from being intercepted or exploited by
malicious actors, especially in the case of information relating to
dangerous dual-use capabilities. The high-level security practices that
developers have been advised to adopt internally (e.g.,
\citep{schuett2023}) should apply to external information disclosures
as well.

A further issue concerns the circumstances of disclosure, that is, the
stages of design, development, and deployment in which developers report
the above information. For example, should the circumstances of
disclosure be determined by reference to a particular time (e.g.
$n$ days or weeks prior to and/or following deployment) or to a
certain capabilities or risk threshold (which may be harder to define)?
Should the timing and frequency of disclosure differ between reporting
information to government actors compared with reporting information to
other developers or domain experts? To operationalize responsible
reporting, we need sufficiently flexible, yet clear, answers to these
questions.

\section{Implementation}\label{sec:5-implementation}

\subsection{Challenges}\label{challenges}

There are several challenges to successfully implementing the proposed
framework for responsible reporting. We divide these challenges into two
categories: (a) challenges facing developers that seek to participate in
the reporting framework and (b) broader challenges concerning the
overall effectiveness of the framework.

\emph{\textbf{Challenges for developers}}. Developers seeking to engage
in responsible reporting are likely to confront four main challenges.

\begin{enumerate}
\item
  \textbf{Intellectual property}. Commercially sensitive information
  (e.g., descriptions of new models or capabilities, logs of real-world
  incidents) could be inadvertently disclosed to, or exploited by, other
  developers and competitors.
\item
  \textbf{Reputational risk}. Reporting safety incidents and anticipated
  risks or vulnerabilities could damage a developer's reputation and
  harm their business interests.
\item
  \textbf{Legal liability}. Disclosing certain safety information could
  potentially increase developers' exposure to legal liability.
  Moreover, a legal obligation to disclose the results of safety tests
  may deter some developers from conducting rigorous safety tests in the
  first place.
\item
  \textbf{Coordination among developers}. Developers may be reluctant to
  participate in the framework and be exposed to business risks without
  assurances that their competitors will also participate and be
  similarly exposed to such risks.
\end{enumerate}

\emph{\textbf{Broader challenges}}. The potential obstacles to the
framework achieving its goals (set out in \Cref{sec:2-goals-of-reporting}) can be grouped into
four broad challenges.

\begin{enumerate}
\item
  \textbf{Evaluation, documentation, and reporting resources}.
  Developers may lack the resources to effectively collect, document,
  and report the information required by the proposed framework.
\item
  \textbf{Misreporting}. Developers may inadvertently or deliberately
  report information that is either inaccurate or incomplete,
  undermining its reliability and usefulness.
\item
  \textbf{Information hazards}. Information reported under the framework
  could be intercepted by malicious actors and used for nefarious
  purposes, or be misused by its intended recipients.
\item
  \textbf{Institutional capacity}. Actors that receive information under
  the framework may lack the capacity to protect, analyze, or
  effectively respond to the information provided.
\end{enumerate}

For further discussion of these and other challenges facing disclosure
mechanisms, see \citep{guha2023}, including concerns relating to
firm-level and broader compliance costs, the impact of disclosure on
design choices in AI development, and the potential for disclosures
adversely impacting governance decisions.

\subsection{Pathways forward}\label{pathways-forward}

Some of the above challenges could be addressed through targeted
institutional mechanisms, some of which are already incorporated in the
proposed framework. Other challenges require broader structural
intervention. In this section, we assess how to address the most salient
concerns along two different pathways.

\vfill\null
\newpage

\subsection*{A. Voluntary
implementation}\label{a.-voluntary-implementation}

The first pathway involves implementing responsible reporting as part of
a voluntary governance regime \citep{bommasani2023, marchant2022,
mcgregor2021}. Developers voluntarily commit to partake in the
reporting framework, whether in the absence of, or alongside, a broader
regulatory regime. In this scenario, we propose the following
institutional mechanisms:

\begin{enumerate}
\item
  \textbf{Differential disclosure}. Concerns regarding the protection of
  intellectual property and commercially sensitive information are
  largely addressed by features of the framework already discussed.
  Developers disclose information pertaining to model development and
  deployment to government actors only, not to competitors or other
  developers. For example, a safety incident report would describe the
  hazardous use observed, but would not disclose the relevant model's
  architecture, training methods, compute, or data. In addition,
  developers could differentially disclose information \emph{within}
  government. For example, developers might disclose information
  about dangerous dual-use capabilities to some (rather than all)
  participating government actors.
\item
  \textbf{Anonymized reporting}. To protect developers' reputations,
  certain potentially damaging information disclosed under the framework
  could be de-identified, such that it could not be attributed to a
  particular developer and would not tarnish their reputation
  \citep{brundage2020, turri2023}. Notably, effective anonymization
  may be difficult to achieve in some circumstances, such as where the
  developer identity can be inferred from the evaluations conducted.
  Anonymization is probably more appropriate for reporting information
  to participating developers and domain experts, not government actors.
  Government actors that demonstrate the characteristics set out above
  (\Cref{recipients}), including reliable information security, should receive
  de-anonymized versions of the information disclosed under the
  framework.
\item
  \textbf{Organizational pre-commitments}. Developers could collectively
  commit in advance to participate in the reporting framework. Such
  commitments could be supported by a bond-like regime in which
  developers make upfront payments (prior to joining the framework) that
  are incrementally returned to developers contingent on their good
  faith participation in the framework.
\end{enumerate}

\subsection*{B. Regulatory
implementation}\label{b.-regulatory-implementation}

The second pathway involves integrating responsible reporting into a
broader purpose-built regulatory regime, such as the U.S. executive
order \citep{thewhitehouseExecutiveOrderSafe2023} or EU AI Act
\citep{europeanparliamenta}. In this scenario, we suggest the
following institutional mechanisms will help facilitate more informative
and actionable reporting:

\begin{enumerate}
\item
  \textbf{Liability safe harbors}. Developers' reluctance to disclose
  information that may increase their legal exposure could be tackled by
  regulation that introduces safe harbor provisions that protect
  companies from legal liability arising from participation in the reporting framework \citep{aipolicyandgovernanceworkinggroup2023,longpre2024}.
  These could be modeled on existing safe harbors in environmental regulation and financial regulation.
\item
  \textbf{Government resourcing}. Under a purpose-built regulatory
  regime, government actors could be allocated resources to develop the
  technical and governance capacity to protect, analyze, and effectively
  respond to information disclosed under the framework
  (e.g., \citep{departmentforscienceinnovation&technology23}).
  Equipped with these resources, government actors could also assist
  developers in making the required disclosures.
\item
  \textbf{Enforcement}. If regulations imposed legal sanctions in the
  event of negligent or deliberate misreporting, developers would be
  strongly incentivized to establish organizational processes for
  ensuring good faith and effective reporting. Independent auditors
  approved by regulators could also assist in detecting misreporting
  \citep{casper2024, falco2021, heim2024, mokander2023}.
\end{enumerate}

\vfill\null
\newpage

\section{Conclusion}\label{sec:6-conclusion}

Improvements in AI safety and governance hinge on the information
available to key stakeholders. Building on existing efforts in
government, industry, and civil society, responsible reporting aims to
facilitate communicating and responding to safety-critical information
in a dedicated secure institutional framework. While the implementation
of responsible reporting faces several challenges, there are promising
pathways forward. Frontier developers could begin by voluntarily
reporting information about risks and mitigations that goes beyond
current regulatory requirements. Policymakers, meanwhile, could
integrate features of responsible reporting into emerging governance
regimes.

\section*{Acknowledgements}\label{acknowledgements}

For helpful comments and suggestions, we thank Conor Griffin, Lewis Ho,
Séb Krier, Lucy Lim, Aalok Mehta, Nikhil Mulani, Cassidy Nelson, Cullen
O'Keefe, Sophie Rose, Yonadav Shavit, and Toby Shevlane.

\clearpage
\appendix

\section*{\Large Appendices}

\section{Cybersecurity}\label{App:A-Cybersecurity}

\vspace{-1mm}
{%
\centering
\setlength\LTcapwidth{\linewidth}
\renewcommand{\arraystretch}{1.5}
\arrayrulecolor{head!30!black}
\begin{longtable}{|p{0.19\linewidth}|p{0.745\linewidth}|}
\caption{\normalsize Examples of cybersecurity information (including cyber capabilities and security vulnerabilities) that developers could report as part of the proposed framework}
\label{tab:ApendixA}\\

\hline\rowcolor{head}\bfseries\centering\arraybackslash Category & \bfseries\centering\arraybackslash Key items \\\hline
\endfirsthead

\multicolumn{2}{c}%
{{\bfseries \tablename\ \thetable{} -- continued from previous page}} \\
\hline\rowcolor{head}\bfseries\centering\arraybackslash Category & \bfseries\centering\arraybackslash Key items \\\hline
\endhead

\hline \multicolumn{2}{|r|}{{Continued on next page}} \\ \hline
\endfoot

\hline \hline
\endlastfoot

\begin{tabular}[t]{@{}>{\centering\arraybackslash}p{\linewidth}@{}}\bfseries Development and deployment\\[0.76cm] {\large\faUsers}\enskip\bfseries Recipients:\\[-5pt] Government actors only\end{tabular} & \begin{tabular}[t]{@{}p{\linewidth}@{}}
\vspace{-11pt}
\begin{enumerate}[leftmargin=*,topsep=0pt,parsep=0pt,partopsep=0pt,itemsep=3pt]
    \item List of training datasets and descriptions of data sanitization or anonymization practices.
    \item List of training-related hardware (including processing units, networking hardware, and peripheral equipment).
    \item Software bill of material (SBOM) for training runs.
    \item List of known vulnerabilities in software/hardware components, products, and libraries used in model development and/or deployment.
    \item List of mitigations and/or justification for using products despite known vulnerabilities.
    \item List of cloud providers, resources, and physical data center locations for training runs.
    \item Is the model intended for use in offensive cyber operations and/or defensive cybersecurity activities?
    \item Is the model intended for use in software, firmware, hardware, or cryptographic development?\vspace{-3mm}
\end{enumerate}
\end{tabular} \\\hline
\begin{tabular}[t]{@{}>{\centering\arraybackslash}p{\linewidth}@{}}\textbf{Risks and harms}\\[0.38cm] {\large\faUsers}\enskip\bfseries Recipients:\\[-5pt] \mbox{Government actors}, developers, and independent domain experts\end{tabular} & \begin{tabular}[t]{@{}p{\linewidth}@{}}
\vspace{-11pt}
    \begin{enumerate}[leftmargin=*,topsep=0pt,parsep=0pt,partopsep=0pt,itemsep=3pt]
    \item Results of: (a) external and internal penetration tests, safety evaluations, and vulnerability scans; (b) bounty programs and disclosed vulnerabilities/exploits; (c) threat modeling assessments.
    \item Description of incidents in which a system: (a) discovered novel vulnerabilities in software, firmware, hardware, or cryptographic products; (b) developed malware utilized in illicit activities; (c) was used to illegally access private data, gain access to an unauthorized network, or exfiltrate sensitive information from a device or network.
    \item Complete incident reports for: (a) external breaches and unauthorized access of model weights or other training infrastructure; (b) insider or third party leaking of model weights or other training infrastructure.
    \item Evidence of a system exhibiting the ability to perform: (a) vulnerability and exploit discovery (including static/dynamic code analysis, protocol reverse-analysis, and vulnerability to exploit conversion); (b) malware development and deployment; (c) social engineering attacks (e.g., phishing); (d) compromise of cryptographic systems or protocols; (e) model self-replication.
    \item Report of cybersecurity-related capability evaluations.\vspace{-3mm}
    \end{enumerate}
\end{tabular} \\\hline
\begin{tabular}[t]{@{}>{\centering\arraybackslash}p{\linewidth}@{}}\textbf{Mitigations}\\[0.38cm] {\large\faUsers}\enskip\bfseries Recipients:\\[-5pt] \mbox{Government actors}, developers, and independent domain experts\end{tabular} & \begin{tabular}[t]{@{}p{\linewidth}@{}}
\vspace{-13.5pt}
    \begin{enumerate}[leftmargin=*,topsep=2pt,parsep=0pt,partopsep=0pt,itemsep=2pt]
    \item Description of security standards, personnel screening, and auditing procedures, including: (a) encryption standards for data at rest and in transit; (b) digital access controls (e.g., RBAC); (c) physical access controls; (d) patch management; (e) versioning controls; (f) backup integrity and verification.
    \item Technical description of how users can interact with the model, including: (a) API specifications, security standards, and auditing practices;\qquad\quad (b) credential provisioning and security practices.\vspace{-0.85em}
\end{enumerate}
\end{tabular}\\\hline
\end{longtable}%
}
\vfill\null\clearpage
\vspace*{-2cm}
\section{Biosecurity}\label{App:B-Biosecurity}
\vspace*{-2mm}
{%
\centering
\setlength\LTcapwidth{\linewidth}
\renewcommand{\arraystretch}{1.5}
\arrayrulecolor{head!30!black}
\begin{longtable}{|p{0.19\linewidth}|p{0.78\linewidth}|}
\caption{\normalsize Examples of biosecurity information that developers could report as part of the proposed framework}
\label{tab:ApendixB}\\

\hline\rowcolor{head}\bfseries\centering\arraybackslash Category & \bfseries\centering\arraybackslash Key items \\\hline
\endfirsthead

\multicolumn{2}{c}%
{{\bfseries \tablename\ \thetable{} -- continued from previous page}} \\
\hline\rowcolor{head}\bfseries\centering\arraybackslash Category & \bfseries\centering\arraybackslash Key items \\\hline
\endhead

\hline \multicolumn{2}{|r|}{{Continued on next page}} \\ \hline
\endfoot

\hline \hline
\endlastfoot

\begin{tabular}[t]{@{}>{\centering\arraybackslash}p{\linewidth}@{}}\bfseries Development and deployment\\[0.76cm] {\large\faUsers}\enskip\bfseries Recipients:\\[-5pt] \mbox{Government actors} only\end{tabular} & \begin{tabular}[t]{@{}p{\linewidth}@{}}
\vspace{-11pt}
\begin{enumerate}[leftmargin=*,topsep=0pt,parsep=0pt,partopsep=0pt,itemsep=3pt]
    \item List of all biology-related data used in training, including papers, experimental protocols, and datasets relating to any life sciences field.
    \item Description of methods to optimize for particular biological capabilities (e.g., host-pathogen interaction prediction, genetic sequence analysis or assembly, or structural outputs) or call specific biological tools.
    \item Evaluation of biological science capabilities (such as conceptual ideation, experimental design, knowledge pooling and teaching, laboratory standard operating procedures and tacit knowledge, and sequence design capabilities).
    \item Description of intended user base and deployment strategy (e.g., laypeople or a particular research or practitioner community, API design and deployment).\vspace{-3mm}
\end{enumerate}
\end{tabular} \\\hline
\begin{tabular}[t]{@{}>{\centering\arraybackslash}p{\linewidth}@{}}\textbf{Risks and harms}\\[0.38cm] {\large\faUsers}\enskip\bfseries Recipients:\\[-5pt] \mbox{Government actors}, developers, and independent domain experts\end{tabular} & \begin{tabular}[t]{@{}p{\linewidth}@{}}
\vspace{-11pt}
    \begin{enumerate}[leftmargin=*,topsep=0pt,parsep=0pt,partopsep=0pt,itemsep=3pt]
    \item Description of biorisk capability evaluations and red-teaming, including: (a) evaluation methodologies and information hazard risk mitigations; (b) team size, types of participants (including independent domain experts), and skillsets (including security mindset); (c) details of scaffolding and finetuning methods (including specialized biology or chemistry tools).
    \item Results of biorisk capability evaluations, especially evidence of dual-use capabilities and marginal improvements of AI models over existing (non-AI) methods, including the ability to:
    (a) provide dual-use biological information; (b) describe which biological agents or constructs are most hazardous and accessible; (c)~instruct how to acquire or synthesize a controlled agent or a pandemic pathogen; (d) perform end-to-end synthesis of a controlled agent; (e) create a viable alternative structure for a controlled agent or enhancements mapping onto experiments of concern; (f) ideate novel biological tools by combining concepts or natural capabilities; (g) assist in the weaponization of biology; (h) exhibit evidence of security mindset with respect to biological vulnerabilities.
    \item Records of the use of biorisk capabilities, including: (a) access to dual-use biorisk capabilities; (b) violations of model usage policies.\vspace{-2.5mm}
\end{enumerate}
\end{tabular} \\\hline
\begin{tabular}[t]{@{}>{\centering\arraybackslash}p{\linewidth}@{}}\textbf{Mitigations}\\[0.38cm] {\large\faUsers}\enskip\bfseries Recipients:\\[-5pt] \mbox{Government actors}, developers, and independent domain experts\end{tabular} & \begin{tabular}[t]{@{}p{\linewidth}@{}}
\vspace{-11pt}
    \begin{enumerate}[leftmargin=*,topsep=0pt,parsep=0pt,partopsep=0pt,itemsep=3pt]
    \item Description of measures to reduce biorisk capabilities and harmful outputs, including: (a) decisions regarding the inclusion (or non-inclusion) in training data of papers, experimental protocols, and datasets relating to dual-use biology; (b) finetuning and other methods that can cause a model to refrain from performing certain biorisk-related tasks (e.g., accessing and delivering controlled agents and potential pandemic pathogens); (c) preventing jailbreaks and other adversarial attacks; (d) decisions regarding the ability (or inability) of models to call third-party biological tools that have not been evaluated for dangerous biological capabilities during assessment.
    \item Monitoring and controlling model usage: (a) collecting know-your-customer (KYC) information on users who seek to access certain dual-use biorisk capabilities (e.g., developing novel pathogens); (b) restricting access to these capabilities, including plans to adhere to the principle of least privilege (PoLP) through user access controls.
    \item Establishment of biosecurity incident, threat alert, and escalation processes, including: (a) investigation procedures in the event users violate usage policies; (b) incident reporting mechanisms for unanticipated post-deployment demonstration of dangerous biological capabilities; (c) response plan to deliberate malicious use of advanced biological capabilities.\vspace{-2.5mm}
\end{enumerate}
\end{tabular}\\\hline
\end{longtable}%
}
\clearpage
\printbibliography

\end{document}